\documentclass[11pt]{article}
\usepackage{epsfig}
\font\tenimbf=cmmib10 at 10pt
\font\sevenimbf=cmmib10 at 6pt
\font\fiveimbf=cmmib10 at 4pt
\newfam\imbf
\textfont\imbf=\tenimbf
\scriptfont\imbf=\sevenimbf
\scriptscriptfont\imbf=\fiveimbf

\begin{document}

\begin{titlepage}
\title{
\vskip .5cm
\vspace{20mm}
\bf{Photon emission in a hot QCD plasma{\footnote {Talk given at 
Strong and Electroweak Matter 2004
Helsinki, Finland, June 2004.}}}
}
\author{P.~Aurenche} 
\maketitle
\begin{center}
\vskip -1.cm
Laboratoire d'Annecy-le-Vieux de Physique Th\'eorique LAPTH{\footnote {UMR 5108 du CNRS, 
associ\'ee \`a l'Universit\'e de Savoie}},\\ 
BP110, F-74941, Annecy-le-Vieux Cedex, France
\end{center}


\vskip 3cm

\begin{abstract}
Various mechanisms of thermal photon production are reviewed and their
implications for heavy ion collisions are briefly sketched.
\end{abstract}
\vfill
\begin{flushright}
LAPTH-Conf-1064/04\\
hep-ph/yymmnn\\
\end{flushright}

\thispagestyle{empty}
\end{titlepage}

\section{Introduction}

It is often mentioned that photons, or more generally electromagnetic probes,
can be used to signal the formation of quark-gluon plasma (QGP) in
ultra-relativistic heavy ion collisions.  They could also be used to measure the
temperature of this plasma at the time of its formation. Data relevant for this
problem will soon cover a wide energy range from a nucleon-nucleon
center-of-mass energy about 20 GeV (Pb-Pb collisions at CERN) to 200 Gev (Au-Au
collisions at RHIC, Brookhaven) and  5.5 TeV (Pb-Pb collisions at LHC after
2007).

Many production mechanisms of electromagnetic probes exist in heavy ion
collisions. On a very qualitative level, one distinguishes the emission of
photons  in the initial stage of the collision (primary production): it is
similar to the case of proton-proton collisions and goes via QCD Compton
scattering, quark-antiquark annihilation and bremsstrahlung emission, involving
partons in the incoming hadrons. It is calculated in perturbative QCD to the
next-to-leading order (NLO). The energy spectrum is power behaved and it is
expected to dominate the high momentum region. Hadrons such as $\pi^0$ and
$\eta$ are also produced by perturbative mechanisms and they decay into
photons  which contribute an important background to the photon signal at
large momentum.
In the heavy ion collision many partons are produced which eventually
thermalise into a quark-gluon plasma. This hot QGP bubble  expands and cools
until a hadronic phase is formed. Photons are produced in the QGP phase as well
as in the hot hadronic phase with a rate which is expected to be exponentially
decreasing with the energy of the photon (secondary production). This thermal
contribution is expected to be important mainly in the ``intermediate" energy
region region up to 10 GeV, say. Thermally produced $\pi^0$'s and $\eta$'s again
contribute an important background.

In the following we review recent progress in the calculation of photon rates
in the QGP phase and present phenomenological applications for RHIC and LHC
heavy ion collisions.

\section{Thermal production of hard photons}

We assume in this section that the plasma is in chemical and thermal
equilibrium  at temperature $T$. The rate of production of a real photon of
momentum $P = (E, \bf p)$, per
unit time and volume, is 
\begin{equation}
 \frac{E \ dN}{dt d{\bf x} d{\bf p}}=-
  \frac{1}{(2\pi)^3}\;n_{_{B}}(E)\,
  {\rm Im}\,\Pi_\mu{}^\mu(E + i \epsilon,{\bf p})\; ,
  \label{realphot}
\end{equation}
where $\Pi_\mu{}^\mu(E + i\epsilon,{\bf p})$ is the retarded photon
polarisation tensor. The Bose-Einstein factor $n_{_{B}}(E)$ provides an
exponential damping when $E\gg T$. A similar formula holds for lepton pair
production.

The two-point correlation function is calculated in the framework of the hard
thermal loop (HTL) resummed theory of Braaten and Pisarski~\cite{BraatP1}. In
this approach one distinguishes two scales: the ``hard" scale, typically of
order $T$ or larger (the energy of quarks and gluons in the plasma) and the
``soft" scale of order $g T$ where $g$, the strong coupling, is assumed to be
small. Collective effects in the plasma  modify the physics at scale $g T$,
{\em i.e.} over long distances of ${O}(1/gT)$. Propagators and vertices are
modified by the summation of higher order diagrams. This is easily illustrated
with the example of the fermion propagator, $S(Q)$, which in the ``bare" theory
is simply $1/p$ (we neglect spin complications and make only a dimensional
analysis)\footnote
	{A low case letter denotes the modulus of the 3-momentum.}.
The thermal contribution to the one loop correction is found to be 
$\Sigma(Q)\sim g^2 T^2 /q$ which is of $O(gT)$ when $q \sim gT$. The resummed
propagator $^*S(Q) = 1/ (q -\Sigma(Q))$ is then deeply modified, compared to
the bare propagator, for soft momenta of ${O}(gT)$ whereas thermal
corrections appear essentially as higher order effects for hard momenta.
Thermal resummation affects the propagator in two ways: 1) in the
space-like region ($Q^2 <0$), the propagator acquires an imaginary part, due to
Landau damping (a new feature compared to the $T=0$ theory) characterised by a
thermal mass scale; 2) in the time-like region a thermal mass, of asymptotic
value  $m^2_{\rm q} = g^2 T^2 C_F/4$, is generated. In both cases thermal
effects screen potential soft or collinear singularities.  Likewise, the gluon
propagator is modified: Debye screening mass $m^2_{_{\rm D}}= g^2
T^2(N_c+N_f/2)/3$  in the space-like region and quasi-particle masses in the
time-like region.\footnote{
	$N_c$ is the number of colors, $N_f$ the number of flavors and $C_F$ the
	Casimir of the fundamental representation.}
One-loop corrections also modify the vertices when the external momenta are
soft~\cite{BraatP1}. One can construct an effective Lagrangian~\cite{BraatP4}
in terms of effective propagators and vertices and calculate observables in
perturbation theory.

\vspace{.5cm}
\begin{figure}[htbp]
\begin{center}
\resizebox*{11cm}{!}{\includegraphics{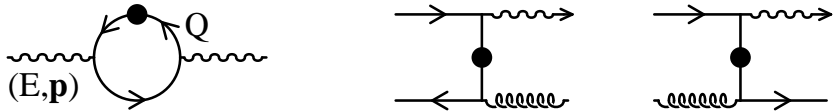}}
\end{center}
\caption{
Left diagram: one-loop contribution; right diagrams: annihilation and Compton
scattering  processes obtained when cutting the one-loop diagram.}
\label{figure1}
\end{figure}
In the one-loop approximation, the production rate of hard photons is given by the
diagram shown in Fig.~\ref{figure1} where the symbol $\bullet$ means that
effective propagators are used. For hard photon momentum it is indeed enough to
consider bare vertices and only one effective fermion propagator, since one
propagator is necessarily hard and needs no resummation. The result is (for one quark
flavor)~\cite{KapusLS1} 
\begin{equation} 
{\rm Im}\,\Pi_{\mu}{}^{\mu}(E,{\bf p}) = \frac{5}{9}\ \alpha\ g^2\  T^2\
\Big(\ln (\frac{E T}{m_{\rm q}^2}) + constant \Big) \, .
\label{eq:1loop}
\end{equation} 
One notices the logaritmic growth of the 2-point function with the energy of
the photon while the thermal quark mass acts as a soft cut-off which screens
the singularity in the forward scattering of the processes in
Fig.~\ref{figure1}.

\begin{figure}[htbp]
\begin{center}
\resizebox*{11cm}{!}{\includegraphics{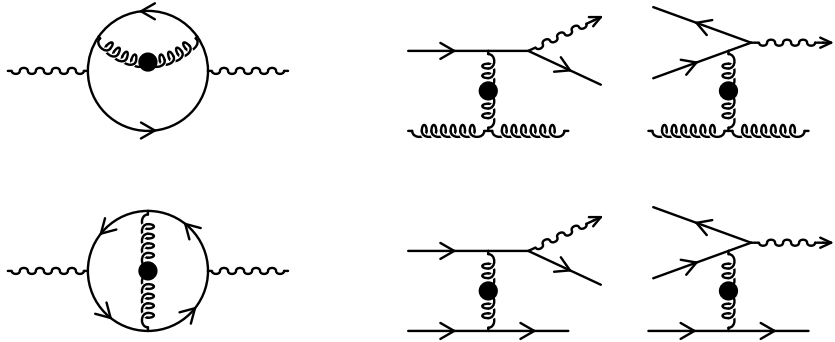}}
\end{center}
\caption{
Left diagrams: two-loop contributions; right diagrams: bremsstrahlung and
off-shell annihilation processes obtained when cutting through the two-loop
diagrams.}
\label{figure2}
\end{figure}
At this point, the HTL picture of photon production, dominated by Compton and
annihilation scattering, turned out to be at variance with the result of a
semi-classical calculation where it was found that the production was dominated
by the bremsstrahlung radiation of photons from quarks~\cite{CleymGR2}. The way
out of this paradox is to consider the two-loop diagrams in the effective
theory (Fig.~\ref{figure2}). A dimensional analysis reveals that the dominant
contribution comes from the kinematical region where the quark momenta in the
loop are hard but the gluon is soft and space-like. The physical processes
exhibited by cutting through the loop diagrams of Fig.~\ref{figure2} correspond
to the bremsstrahlung emission from a quark or an anti-quark scattering in the
plasma (as in the semi-classical approximation) and by a new process, the
off-shell annihilation of a quark-antiquark pair where one of the (anti-)quark
scatters in the plasma. For two flavors, the result of the calculation is very
simple~\cite{us1}: 
\begin{equation}
{\rm Im}\,\Pi_\mu{}^\mu(E,{\bf p}) = \frac{5}{9}\ \alpha \ g^2\ \frac{8}{3}\  
\left[\frac{T^3}{E} + \frac{E T}{\pi^2} \right]\, ,
\label{eq:2loop}
\end{equation}
where the first term arises from bremsstrahlung and the second one from
off-shell annihilation. In the general case the numerical coefficient is replaced by a function of the
ratio of the Debye mass over the thermal quark mass~\cite{us1}. The two-loop HTL
result is not suppressed, as expected, by powers of the coupling $g$ compared
to the one-loop result. This is due to an enhancement factor of
$O(T^2/m^2_{\rm q}) \sim 1/g^2$ associated to the collinear emission of the
photon.

One may wonder about the usefulness of the loop expansion in the HTL effective
theory if the two-loop rate is of the same order as the one-loop rate. It turns
out that higher loop diagrams contribute to leading order~\cite{us2}. The
physical reason is the following. Consider a virtual quark of momentum $R$ 
decaying into a photon of momentum $P$ and a quark of momentum $Q$. The photon
formation time ({\em i.e.} the life-time of the virtual state) is  $\tau_{_{\rm
F}} \equiv r / {R^2} \equiv q r /  (p ({\bf q}^2_{_T} + m^2_{\rm q}))$  with
the transverse momentum of the quark measured with respect to the photon
momentum. This formation time is long, of $O(T/m^2_{\rm q}) \sim 1/g^2 T$, and
is comparable to the mean free path of the quark in the plasma given by the
inverse of the quark damping rate $\gamma$. Quark rescattering in the medium
should thus be taken into account. This is done by resumming the self-energy
diagrams on the quark propagator going beyond the hard thermal loop
approximation and including also the imaginary part (related to $\gamma$)
generated by rescattering. In our case the time-like quark propagator, which in
the HTL theory was $1/(q^2_{_T} +  m^2_{\rm q})$ becomes approximately
$1/(q^2_{_T} +  m^2_{\rm q} + i \gamma q r / p)$: this shows that both
thermal mass and damping rate act as competing cut-offs.
\vspace{.5cm}
\begin{figure}[ht]
\begin{minipage}[c]{7.cm}
\centerline{\includegraphics[width=7.cm]{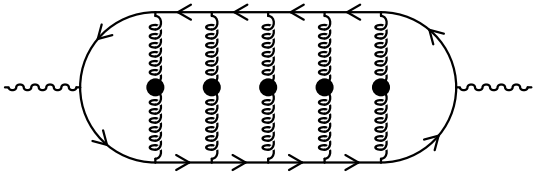} }
\end{minipage}
\hfill
\begin{minipage}[l]{5.cm}
   \caption{
     An example of a ladder diagram.
     }
\label{figure3}
\end{minipage}
\end{figure}
\vspace{.5cm}
Of course, to
respect gauge invariance, one should resum vertex corrections in the same
approximation as self-energy corrections~\cite{ArnolMY1}, which motivates the
fact that ladder diagrams such as the graph of  Fig.~\ref{figure3} should be
summed. Arnold, Moore and Yaffe~\cite{ArnolMY1} showed that such diagrams
indeed contribute to leading order and they derived a very elegant integral
equation to perform the resummation. The imaginary part of the photon
polarisation tensor takes the form:
\begin{eqnarray*}
{\rm Im}\,\Pi_\mu{}^\mu(P)\approx 
{\alpha N_c}
\int_{-\infty}^{+\infty}dq
\,[n_{_{F}}(r)-n_{_{F}}(q)]\;
\frac{q^2+r^2}{q^2 r^2}
\,{\rm Re}\int \frac{d^2{\bf q}_{_T}}{(2\pi)^2}\;
{\bf q}_{_T}\cdot{\bf f}({\bf q}_{_T}) \, , 
\label{eq:int}
\end{eqnarray*}
with $r = p+q$ and $n_{_{F}}(q) = 1/(\exp(q/T)+1)$, the Fermi-Dirac statistical
weight. The dimensionless resummed quark-quark-photon vertex, ${\bf f}({\bf
q}_{_T})$,  is shown by the shaded vertex in Fig.~\ref{figure4} while ${\bf
q}_{_T}$ is proportional to the bare vertex. The value of ${\bf f}({\bf
q}_{_T})$ is obtained by solving the following equation~\cite{ArnolMY1}:
\begin{figure}[ht]
\vspace{.5cm}
\centerline{\epsfxsize=11.5cm\epsfbox{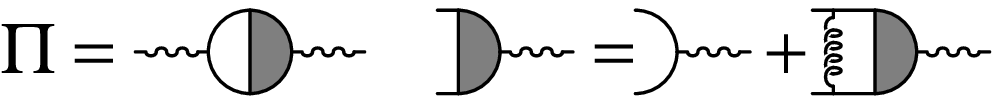}}   
\vspace{.3cm}
\caption{The integral equation resumming the ladder diagrams.}
\label{figure4}
\end{figure}
\begin{equation}
\frac{i}{2 \tau_{_{\rm F}}}{\bf f}({\bf q}_{_T})
= 2{\bf q}_{_T} + g^2 C_{_F} T \!\! \int
\frac{d^2{\bf l}_{_T}}{(2\pi)^2} \, 
{C}({\bf l}_{_T}) \,
[{\bf f}({\bf q}_{_T}+{\bf l}_{_T})-{\bf f}({\bf q}_{_T})]\; ,
\label{eq:integ-f}
\end{equation}
where $\tau_{_{\rm F}}$ is the formation time defined above. The collision
kernel is ${C}({\bf  l}_{_T})={m^2_{_{\rm D}}}/{{\bf l}_{_T}^{\ 2}} ({{\bf 
l}_{_T}^{\ 2}+{ m^2_{_{\rm D}}}})$ and it includes the exchange of both
longitudinal and transverse gluons~\cite{us2}. Solving for this integral
equation one can obtain a fit, in the large $E/T$ range 
\begin{equation}
{\rm Im}\,\Pi_\mu{}^\mu(P)\approx \alpha \ g^2\  T^2\ \sqrt{\frac{E}{T}}\, ,
\end{equation}
to be compared to the linear in $E$ behaviour of Eq.~(\ref{eq:2loop}). This
is an illustration of the Landau-Pomeranchuck-Migdal (LPM) suppression of hard
photons due to multiple scattering of the quark in the plasma. 

Rates for small mass lepton pairs produced at large momentum have also been
calculated at one-loop~\cite{AltheR1}, two-loop~\cite{us3} and including the LPM
effect~\cite{us4}.

\section{Phenomenology at RHIC and LHC}

\begin{figure}[bt!]
\centerline{\epsfxsize=5.5in\epsfbox{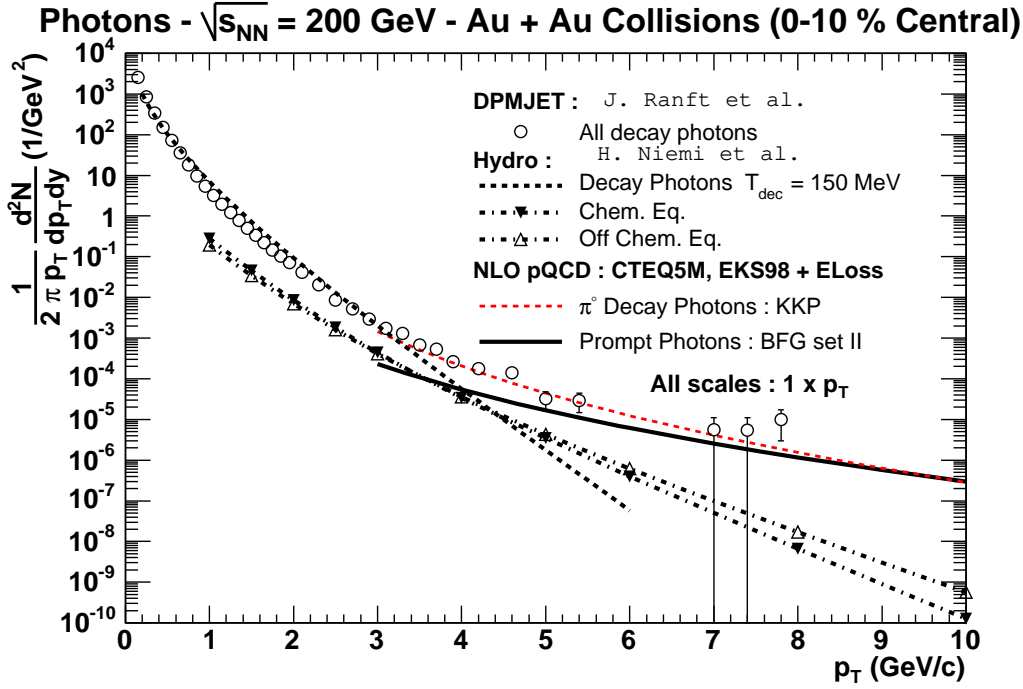}}   
\caption{Various contributions to the production of photons in Au-Au 
collisions at RHIC. The open dots are the predictions of decay photons from the
non-thermal DPMJET model~{\protect\cite{ranft}} for heavy ion collisions, in good
agreement with the hydrodynamic $+$ perturbative QCD model.}
\label{figure5}
\end{figure}
\begin{figure}[htb!]
\centerline{\epsfxsize=5.5in\epsfbox{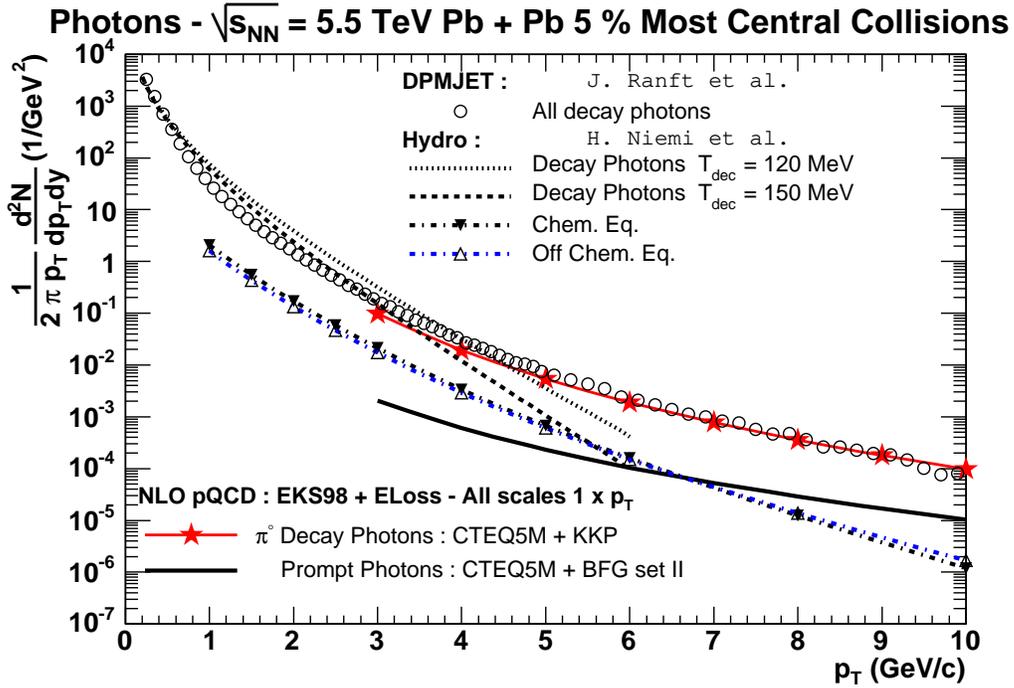}}   
\vspace{-.5cm}
\caption{Same as the above figure for Pb-Pb collisions at LHC.}
\label{figure6}
\end{figure}

To make predictions for photon production in heavy ion collisions, the rates
calculated above have to be included in a hydrodynamical code which describes
the expansion and the cooling of the plasma from chosen initial state
conditions. In the following we use the code of Ruuskanen et
al.~\cite{ruuskanen}. The rate for thermal photon production is shown by the
thick dash-dotted line in Fig.~\ref{figure5} for Au-Au central collisions at RHIC and in
Fig.~\ref{figure6} for Pb-Pb central collisons at LHC. Further details on the
model can be found in Ref.~\cite{yellowreport}. Also shown on the figures are the
rates of production of decay photons from thermally produced resonances
(mainly $\pi^0$ and $\eta$) (thick dashed lines) which dominate in the low
$p_{_T}$ range. We also display the rates of primary photons produced in
perturbative QCD mechanisms (solid lines) contributing in the large $p_{_T}$ range as well as
the decay photons from perturbatively produced resonances. In these
estimates account has been taken of energy loss effects on the jet
fragmentation in the medium using the model of Ref.~\cite{sarcevic}.
\begin{figure}[hb!]
\centerline{\epsfxsize=5.5in\epsfbox{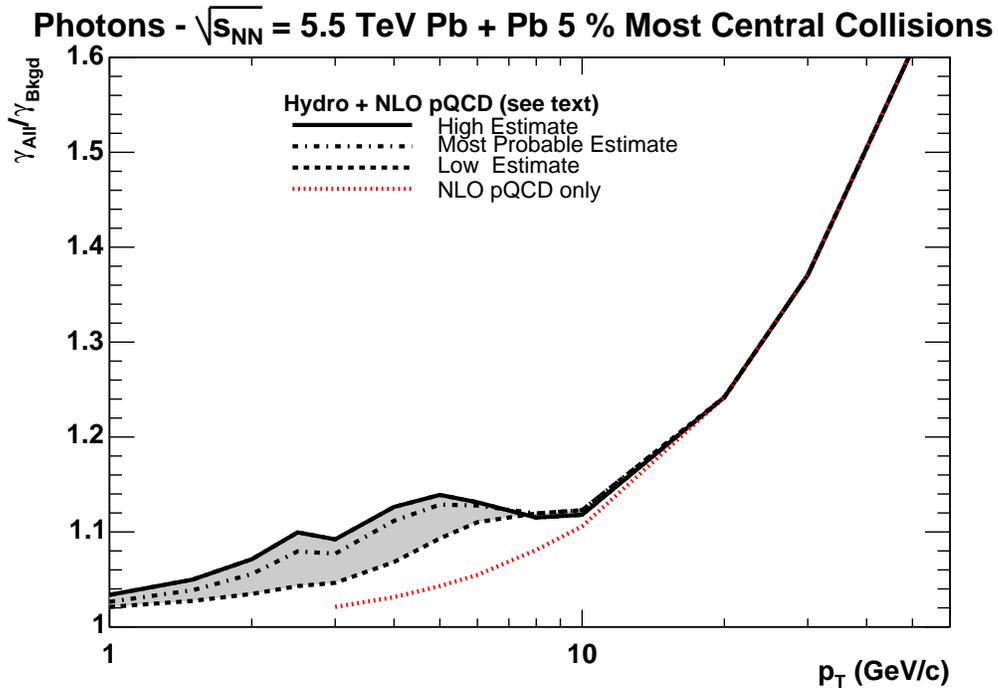}}   
\vspace{-.5cm}
\caption{Ratio of all photons over decay photons at LHC with thermal
production (thick lines) and without thermal production.}
 \label{figure7}
 \end{figure}

It is instructive to look at the ratio of ``all" produced photons over decay
photons and to compare it with the prediction for the same ratio assuming no
thermal production. This plot quantifies the excess of direct photons
produced by the hot hadronic medium formed in a heavy ion collision. At
LHC, thermal effects are important but the ratio of directly produced photon
remains only about 10\%.
\begin{figure}[ht]
\centerline{\epsfxsize=5.5in\epsfbox{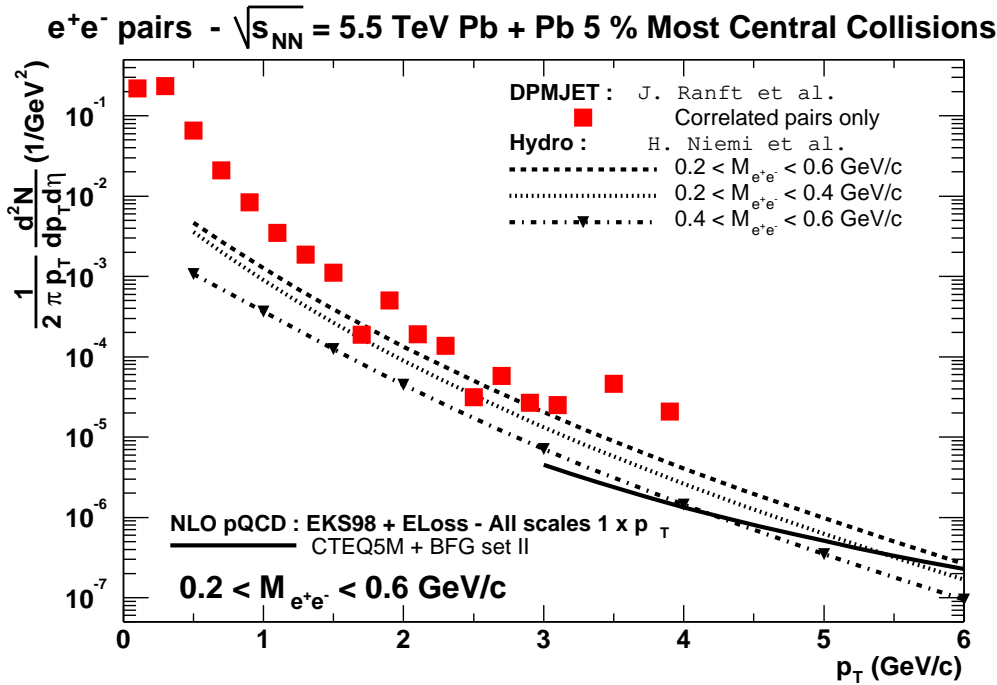}}   
\caption{Thermally produced lepton pairs (dashed line) compared to
perturbatively produced pairs (solid line) and background from hadronic decays
(squares).}
\label{figure8}
\end{figure}

A channel which should suffer less background than direct photon is that of
small mass lepton pair at large momentum: indeed choosing a pair mass above the
$\pi^0$ mass should eliminate a large background. The thermal production rates
have been calculated and included in the hydrodynamical
code~\cite{yellowreport}. The result is shown in Fig.~\ref{figure8}.

\section{Further developments}

In the discussion above we have considered two independent production
processes: emission of the photon in the initial stages of the collision and
production in the quark-gluon plasma and hot hadronic phase. One can envisage
``mixed" production mechanisms where a photon is emitted in the interaction of
a hard quark or gluon (calculable in perturbative QCD) with a quark or the
gluon of the plasma. Since the rate of hard partons is power
behaved while thermal interactions are exponentially damped, the resulting rate
of photon production is expected to be power behaved and to dominate over
purely thermal processes at large momentum. Such examples of ``mixed"
interactions have been considered in Ref.~\cite{Fries:2002kt} who made
simplified estimates of the Compton ($g q \rightarrow \gamma q$) and
annihilation ($q \bar q \rightarrow \gamma g$)  processes when one of the
initial $q$ or $g$ is hard and the other parton is thermalised.
\begin{figure}[htb]
\begin{minipage}[c]{8.cm}
\vspace{.5cm}
\centerline{\includegraphics[width=7.6cm]{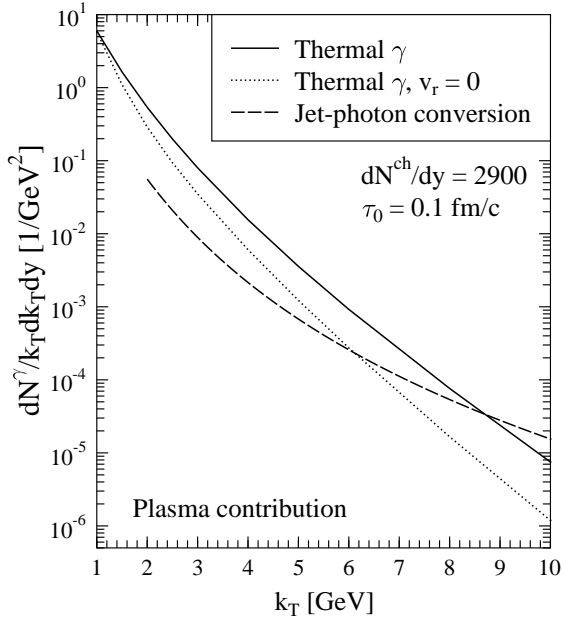} }
\end{minipage}
\hfill
\begin{minipage}[l]{4.cm}
   \caption{
     Photon production from ``mixed" processes: rate from the Compton and annihilation processes between
     a hard parton with a parton in the plasma (dashed line); 
     thermal photon production under the same hypothesis of no transverse
     expansion (dotted line). For comparison the thermal rate with radial
     expansion, used in the previous section, is also shown (solid line).
     }
\label{figure9}
\end{minipage}
\vspace{.5cm}
\end{figure}
 The result is shown in Fig.~\ref{figure9} by the dashed line to be compared
with the dotted line which is the estimate of the purely thermal process under
the same hydrodynamical conditions (no radial expansion). Recently B. Zakharov
estimated the enhanced bremsstrahlung emission  of a photon by a hard quark
traversing the plasma~\cite{zakha}. He found that, due to finite size effects,
the usual $1/x$ spectrum characteristic of bremsstrahlung of a quark in vacuum
is replaced by an approximate $1/(1-x)$ spectrum for  $x<1$. This is
illustrated in Fig.~\ref{figure10} which compares the  fragmentation function
of a quark into a photon in vacuum with the same  in the medium taking into
account only one rescattering in the plasma. The change of shape of the
spectrum will lead to an enhanced production of photon by bremsstrahlung
compared to production in the vacuum. Multiple scattering effects will somewhat
reduce the size of the predicted enhancement~\cite{zakha}. In the figures of
the previous section we already included medium effects on bremsstrahlung
production of photons and this lead to a suppression of photon emission
compared to vacuum which seems to contradict the present result. In fact, there
is no contradiction if we consider that in the latter case the photon was
emitted by the quark outside the plasma while the former case one estimates the
emission of the photon in the plasma.
\begin{figure}[ht]
\centerline{\epsfxsize=5.in\epsfbox{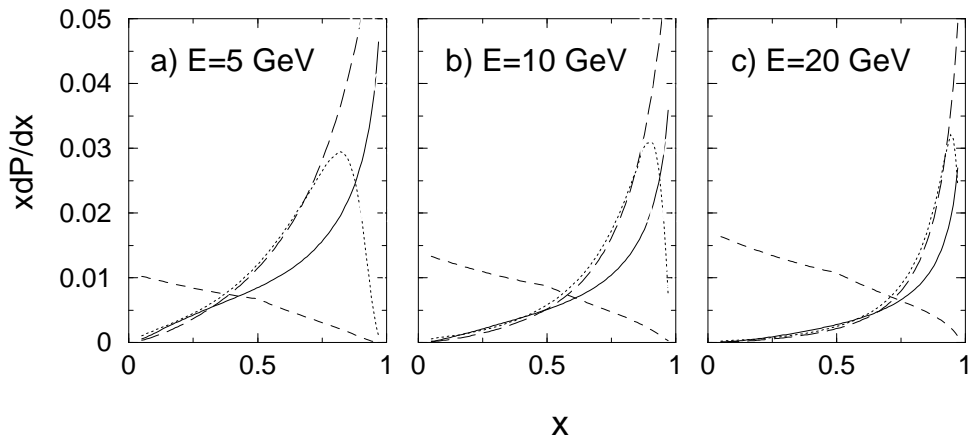}}   
\caption{Comparison of the fragmentation function of a quark into a photon in
vacuum (short dashed line) and in the medium (dotted line) with one scattering
of the hard quark in the plasma. The other curves are for an infinite medium.}
 \label{figure10}
 \end{figure}

It would be important to have a complete and consistent estimate of the mixed
production rates, using a realistic hydrodynamical model: in any case, they
will increase the thermal production rates calculated in the previous section.

All previous results are obtained assuming local thermal equilibrium and the
rates are calculated per unit time. In a series of papers Boyanovsky et
al~\cite{boyadevega} proposed a model to explicitely account for the finite
life-time of the plasma. They assume that photon emission starts at an initial
time $t_i$ and ends at time $t_f$:
under this assumption, usually
forbidden processes of type $q \rightarrow q \gamma$ contribute since energy
conservation does not hold. As a consequence a transient term of type $\alpha
\ln t$, where $t=t_f-t_i$, appears besides the usual $\alpha \alpha_s t$ term,
and it is claimed to dominate the photon production rate at LHC at large enough
momenta. Several objections have been raised concerning the validity of this
picture~\cite{dadic,serreau,fraga}. In particular the hypothesis of a sharp
switching on and off of the interactions is probably not adequate for the
description of the physics of the problem.

\section*{Acknowledgments}
I thank F. Gelis and V. Ruuskanen for collaboration and many discussions.
I also thank F. Gelis for a careful reading of the manuscript.



\begin{thebibliography}{99}


\bibitem{BraatP1}
{E. Braaten, R.D. Pisarski}, {\it Nucl. Phys.}  {\bf B337}, 569 ({1990}); 
{\bf B339}, 310 ({1990});
{J. Frenkel, J.C. Taylor}, {\it Nucl. Phys.}  {\bf B334}, 199 ({1990});
{\bf B374}, 156 ({1992}).

\bibitem{BraatP4}
{J.C. Taylor, S.M.H. Wong}, {\it Nuc. Phys.} {\bf B346}, 115 ({1990});
{E. Braaten, R.D. Pisarski}, {\it Phys. Rev.} {\bf D45}, 1827 ({1992}).

\bibitem{KapusLS1}
{J.I. Kapusta, P. Lichard, D. Seibert}, {\it Phys. Rev.} {\bf D44} ({1991}) 2774;
{R. Baier, H. Nakkagawa, A. Niegawa, K. Redlich}, {\it Z. Phys.} {\bf C53},
({1992}) 433.

\bibitem{CleymGR2}
{J. Cleymans, V.V. Goloviznin, K. Redlich}, {\it Phys. Rev.} {\bf D47}, 989
(1993); {V.V.~Goloviznin, K.~Redlich}, {\it Phys. Lett.} {\bf B} {\bf 319}, 520
({1993}).

\bibitem{us1}
P. Aurenche, F. Gelis, R. Kobes, H. Zaraket, {\it Phys. Rev.}  {\bf D58},
  085003 ({1998});
P. Aurenche, F. Gelis, H. Zaraket, {\it JHEP} {\bf 0205}, 043 (2002).

\bibitem{us2}
P. Aurenche, F. Gelis, H. Zaraket, {\it Phys. Rev.} {\bf D61}, 116001 (2000);
{\bf D62}, 096012 (2000).

\bibitem{ArnolMY1}
{P. Arnold, G.D. Moore, L.G. Yaffe}, {\it JHEP} {\bf 0111}, 057 ({2001});
 {\it JHEP} {\bf 0112}, 009 ({2001});
{\it JHEP} {\bf 0206}, 030 ({2002}).

\bibitem{AltheR1}
{T. Altherr, P.V. Ruuskanen}, {\it Nucl. Phys.} {\bf B380}, 377 ({1992});
{M.H. Thoma, C.T. Traxler}, {\it Phys. Rev.} {\bf D56}, 198 ({1997}). 

\bibitem{us3}
P. Aurenche, F. Gelis, H. Zaraket, {\it JHEP} {\bf 0207}, 063 (2002).

\bibitem{us4}
P. Aurenche, F. Gelis, G.D. Moore, H. Zaraket, {\it JHEP} {\bf 0212}, 006 (2002).

\bibitem{ruuskanen} 
K.J. Eskola, K. Kajantie, P.V. Ruuskanen, {\it  Eur. Phys.J.} {\bf C1}, 627 (1998); 
K.J. Eskola, P.V. Ruuskanen, S.S. R\"as\"anen, K. Tuominen, {\it Nucl. Phys.} {\bf  A696},
715 (2001);
K.J. Eskola, H. Niemi, P.V. Ruuskanen, S.S. R\"as\"anen, {\it Nucl. Phys.} {\bf A715},
561  (2003); 
K.J. Eskola, H. Niemi, P.V. Ruuskanen, 
{\it Phys. Lett}. {\bf B566}, 187 (2003).

\bibitem{ranft}
R.\ Engel and J.\  Ranft, {\it Phys. Rev.} {\bf D54}, 4244 (1996);
S.\ Roesler, R.\ Engel and J.\  Ranft, hep-ph/0012252;
J.\  Ranft, hep-ph/9911213, hep-ph/9911232.

\bibitem{yellowreport}
F.~Arleo et al., 
{\em Photon physics in heavy ion collisions at the LHC},
hep-ph/0311131,
to appear in the CERN Yellow Report on {\em Hard probes in heavy
ion collisions at the LHC}.

\bibitem{sarcevic} J. Jalilian-Marian, K. Orginos and I. Sarcevic,
{\it Phys. Rev.} {\bf C63},  041901 (2001); {\it ibid.}
{\bf A700}, 523 (2002);
S. Jeon, J. Jalilian-Marian and I. Sarcevic,  {\it Phys. Lett.}  {\bf B562},
45 (2003).

\bibitem{Fries:2002kt}
R.~J.~Fries, B.~Muller and D.~K.~Srivastava,
{\it Phys.\ Rev.\ Lett.}\  {\bf 90}, 132301 (2003).

\bibitem{zakha} 
B.G. Zakharov, hep-ph/0405101.

\bibitem{boyadevega} 
D. Boyanovsky, H. J. de Vega, {\it Phys. Rev.} {\bf D68}, 065018 (2003); hep-ph/0311156.

\bibitem{dadic}
I. Dadic, F. Gelis, G.D. Moore, J. Serreau in Ref.~\cite{yellowreport}, Sec. 9.

\bibitem{serreau} 
J.~Serreau, {\it JHEP} {\bf 0405}, 078 (2004).

\bibitem{fraga}
E. Fraga, F. Gelis, D. Schiff, hep-ph/0312222.


\end{thebibliography}
\end{document}